\documentclass{article}
\usepackage{ijcai22}

\usepackage{times}
\usepackage{soul}
\usepackage{url}
\usepackage[hidelinks]{hyperref}
\usepackage[utf8]{inputenc}
\usepackage[small]{caption}
\usepackage{graphicx}
\usepackage{amsmath}
\usepackage{amsthm}
\usepackage{amsfonts}
\usepackage{booktabs}
\usepackage{algorithm}
\usepackage{algorithmic}
\usepackage{subfigure}
\usepackage{bm}
\usepackage{bbm}
\usepackage{color}
\usepackage{array}
\usepackage{multirow}

\newcolumntype{P}[1]{>{\centering\arraybackslash}p{#1}}
\newcolumntype{M}[1]{>{\centering\arraybackslash}m{#1}}
\newcommand{\nosection}[1]{\vspace{2pt}\noindent\textbf{#1.}}

\title{HCFRec: Hash Collaborative Filtering via Normalized Flow with Structural Consensus for Efficient Recommendation}

\author{
Fan Wang
\and
Weiming Liu\and
Chaochao Chen\and
Mengying Zhu\And
Xiaolin Zheng\thanks{Corresponding author}
\affiliations
College of Computer Science and Technology, Zhejiang University, China \\
fanwang1997@hotmail.com,
\{21831010, zjuccc, mengyingzhu, xlzheng\}@zju.edu.cn
}
\begin{document}

\maketitle

\begin{abstract}
    The ever-increasing data scale of user-item interactions makes it challenging for an effective and efficient recommender system. Recently, hash-based collaborative filtering (Hash-CF) approaches employ efficient Hamming distance of learned binary representations of users and items to accelerate recommendations.
However, Hash-CF often faces two challenging problems, i.e., \textit{optimization on discrete representations} and \textit{preserving semantic information in learned representations}.
To address the above two challenges, we propose HCFRec, a novel Hash-CF approach for effective and efficient recommendations. Specifically, HCFRec not only innovatively introduces normalized flow to learn the optimal hash code by efficiently fit a proposed \textit{approximate mixture  multivariate  normal  distribution}, a continuous but approximately discrete distribution, but also deploys a cluster consistency preserving mechanism to preserve the semantic structure in representations for more accurate recommendations. Extensive experiments conducted on six real-world datasets demonstrate the superiority of our HCFRec compared to the state-of-art methods in terms of effectiveness and efficiency.
\end{abstract}

\section{Introduction}

Recommender System (RS) has recently become a crucial tool to alleviate information overload in many areas with rich data, including but not limited to e-commerce, education, finance, and health \cite{Zhu2021A}. As a pivotal technology of RS, Collaborative Filtering (CF) has received thrilling success owing to its inherent domain-independent and easy-to-explain properties. Specifically, CF attempts to learn representations of users and items from their interactive information for the subsequent user preference prediction and item recommendation\cite{chen2022differential}. However, with the explosive growth of users and items, CF based recommendation methods often suffer from high time and space costs \cite{Qi2021Privacy}.

Fortunately, hash-based CF (Hash-CF) approaches \cite{Chen2018Distributed} have been proven to have a good ability to compress data and accelerate computation for recommendations with billion-scale users and items \cite{Shan2018Recurrent}.
In large-scale recommendation scenarios, Hash-CF takes effect by encoding high dimensional real-valued vectors into compact one-hot codes (\textit{hash representations}), such that: (1) bit-wise operations (e.g., XOR) instead of real-valued calculations for preference inference can dramatically accelerate recommendations; (2) bit-wise representations often achieve a $64\times$ storage compression rate compared to real-valued representations. Therefore, Hash-CF enables a lighter and more efficient recommendation model even massive data are involved in decision-makings.

However, existing Hash-CF approaches often face two challenges existing in the learned hash representations.
Firstly, \textbf{CH1:} \textit{how to implement optimization on discrete hash representations?}
The hash representation is usually obtained through the $\rm sign$ function, and optimizing such a representation will lead to a challenging mixed-binary-integer optimization problem \cite{Wang2018a}, which is NP-hard. A promising solution is to replace the $\rm sign$ function with a continuous relaxation (e.g., $\rm tanh$ function) to learn deterministic hash representations in an end-to-end manner, which, however, is not robust due to a lack of noise tolerance consideration. Fortunately, Variational Autoencoder (VAE) \cite{kingma2014} with its probabilistic nature can model features as distributions to accommodate much uncertainty or noisy in data, so as to implement robust recommendation \cite{Liang2018VAE}. However, for the Hash-CF task, features need to be modeled as latent discrete Bernoulli distributions to generate hash representations. Such distributions with discrete nature make the optimization on hash representations more difficult. Secondly, \textbf{CH2:} \textit{how to preserve semantic information in discrete representations?}A hash representation has intrinsically limited representation ability, as it carries less semantic information than a real-valued representation.
Although the existing Hash-CF approaches \cite{Zhang2016DCF} try to control quantization loss to reduce the difference between real-valued and hash representations, they fail to preserve the semantic structure consistency between them. The two representations without structural consensus drop much semantic information that is crucial to accurate recommendations.

In light of the above two challenges, we propose a novel Hash-CF recommendation approach, i.e., HCFRec, which generates compact yet informative hash representations for effective and efficient recommendations. The proposal is comprised of two major components based on VAE framework for robust generalization capability. Specifically, for \textbf{CH1}, \textit{hash representation generation component} first models user (item) features as user (item) real-valued representations that obey a simple prior normal distribution. Considering the difficulty in optimizing a discrete distribution, we innovatively propose an \textit{approximate mixture multivariate  normal  distribution}, a continuous but approximately discrete distribution, and introduce normalized flow \cite{Rezende2015} to deploy reversible transformation functions from the simple normal distribution to the complex and approximately discrete distribution. Normalized flow can efficiently implement reversible distribution transformation from a simple continuous distribution to any complex continuous distribution with same dimensions. To our best knowledge, this is the first attempt to introduce normalized flow for binary optimization. Finally, hash representations are generated by binarization operations.
For \textbf{CH2}, \textit{cluster consistency preserving component} first respectively clusters the real-valued representations and the hash distribution both generated in the first component. Subsequently, a loss function is designed for the two representations to alternately learn semantic information with structural consensus.

We summarize our main contributions as follows: (1) We propose a novel Hash-CF approach implemented on a variational framework for effective and efficient recommendation. (2) Normalized flow is first introduced to learn the optimal hash codes by efficiently fitting our proposed \emph{approximate mixture multivariate normal distribution}, which makes it possible for efficient optimization on discrete distribution. (3) We innovatively develop a cluster structure preserving mechanism to retain cluster consensus between real-valued and hash representations for more accurate recommendations. (4) Extensive experiments conducted on six real-world datasets demonstrate the superiority of our HCFRec approaches over state-of-the-art competitive ones.

\begin{figure*}[t]
\centering
% \vspace{-0.5cm}                         %调整图片与上文的垂直距离
\setlength{\abovecaptionskip}{0.2cm}      %调整图片标题与图距离
\setlength{\belowcaptionskip}{-0.5cm}
\includegraphics[ width=\textwidth]{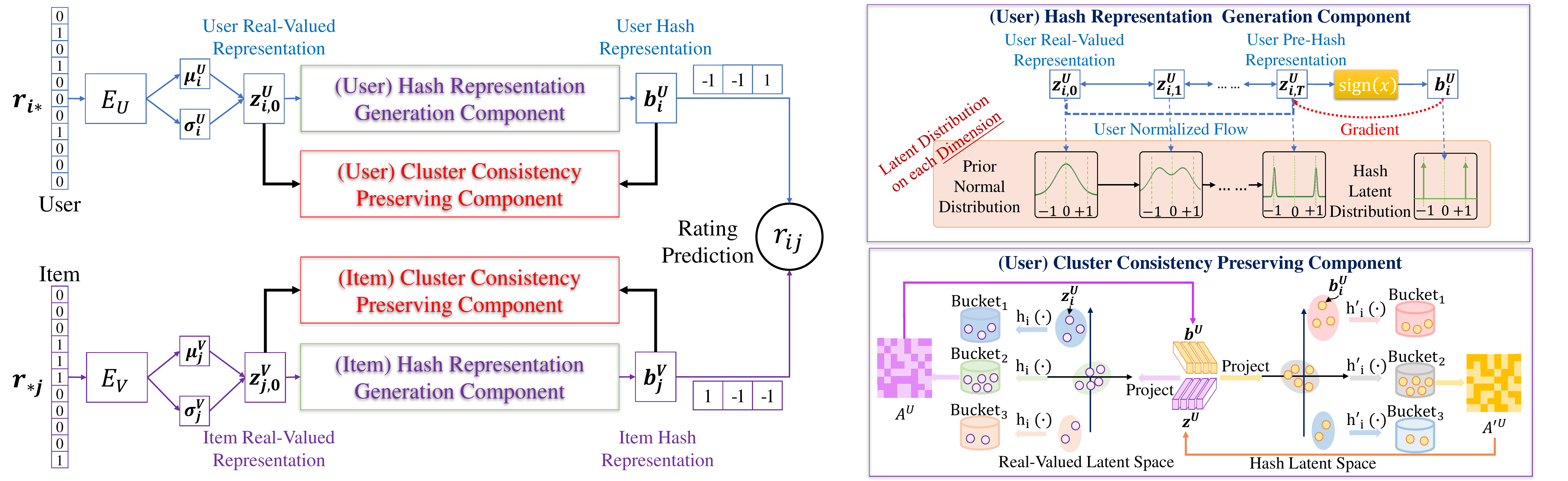}
\caption{The overview of HCFRec. The left is the model framework based on dual VAE, comprised of hash representation generation component and cluster consistency preserving component. The right is the details of the two components, where we employ user side as a specific example.
}
\label{overview}       % Give a unique label
\end{figure*}
\section{Related Work}
\subsection{Hash-CF Methods}
Due to the intrinsic characteristics of high efficiency and low storage cost, Hash-CF has recently gained ever-increasing attentions of researchers, including two kinds of approaches.
First, the "two-stage" approaches learn hash representations through a continuous representation learning stage followed by a binary quantization stage \cite{Karatzoglou2010,Zhang2014Preference,Zhang2017Dot}, which, however, produces considerable quantization errors.
Second, the ``end-to-end" approaches try to optimize hash representations directly. Some researchers model user and item features as deterministic representations \cite{Zhang2016DCF}\cite{Zhang2017DiscretePR}\cite{Liu2019Compositional}. However, these approaches lack noise tolerance. For more generalization, some researchers recruit probability-based VAE to model user/item features as discrete Bernoulli distributions \cite{Hansen2020Content}. However, discrete latent space makes it challenging for hash codes to be optimized effectively and efficiently.

\subsection{VAE-based CF Methods}
VAE has recently achieved strong performance improvements on CF \cite{Liang2018VAE}. VAE’s strong generative ability even in sparse settings is mainly because it can model representations as a distribution rather than a deterministic vector to account for much uncertainty in the latent space. \cite{Lee2017} incorporates auxiliary information into VAE for CF. \cite{Karamanolakis2018} also takes side information into account, which learns user representations in multimodal latent space for better recommendation. RecVAE \cite{Karamanolakis2018} introduces some regularization techniques into VAE to improve recommendation performance. However, despite remarkable improvements are achieve in these works, they only learn user representations but ignore valuable item information. In this situation, \cite{Truong2021} models a Bilateral VAE framework to learn both user and item representations for more robust recommendation.

\section{Methodology}
The data we intend to learn from is a user-item interaction matrix $\bm{R} \in \mathbb{R}^{N_U \times N_V}$, where $N_U$ and $N_V$ denote the number of users and items, respectively. The $i$-th row of $\bm{R}$ is formulated as $\bm{r_{i,*}}\in\mathbb{R}^{N_V}$ while the $j$-th column of $\bm{R}$ is formulated as $\bm{r_{*,j}}\in\mathbb{R}^{N_U}$, representing user feature and item feature, respectively. Moreover, $E_U$ and $E_V$ are respectively user encoder and item encoder, $\bm{b^{U}}\in \{-1,1\}^D$ and $\bm{b^{V}}\in \{-1,1\}^D$ respectively indicate user hash representations and item hash representations. The main problem we need to solve is: \textit{Given $\bm{r_{i,*}}$ and $\bm{r_{*,j}}$ from matrix $\bm{R}$, a model is needed to learn hash representations $\bm{b^{U}}$ and $\bm{b^{V}}$ for effective and efficient recommendations.}
\subsection{Overview of HCFRec}
Our HCFRec model is mainly based on a dual VAE framework comprised of \textit{hash representation generation component} and \textit{cluster consistency preserving component}, as illustrated in Figure \ref{overview}. Details of each component are as follows:\\
\nosection{(1) Hash Representation Generation Component} Due to the close correlation with the VAE framework, this component is embedded into the VAE for introduction here. Concretely, we first employ VAE framework to model user (item) features as real-valued representations that obey prior normal distribution. Normalized flow is subsequently deployed to achieve pre-hash representation that obeys a continuous but approximately discrete distribution. Finally, $\rm sign$ function on this distribution is recruited to output hash representations.

\nosection{(2) Cluster Consistency Preserving Component} We cluster the embedded real-valued representations and the generated hash representations, and alternately learn cluster-consistent semantic information from each other to ensure the semantic structure  restoration ability of hash representations. Finally, HCFRec uses the enhanced hash representations for recommendations.

%We elaborate the two components as follows.

%We first employs encoder $E_U$ ($E_V$) in VAE to model user (item) features as user (item) real-valued representation that obey a simple prior normal distribution. Then, the normalized flow deploys reversible distribution transformation function to achieve user (item) pre-hash representation that obeys an continuous but approximately discrete distribution. Finally, sign($\cdot$) on this distribution can be recruited to output the user (item) hash representations. \\

\subsection{Hash Representation Generation Component}
\label{3.1}
\nosection{Modelling} For a recommendation task, the objective we seek to maximize is the likelihood of a rating, i.e., $p(r_{i,j})$. Formally, we determine the likelihood $p(r_{i,j})$ in Eq.(\ref{rating_likelihood}) by introducing binary latent representations $\mathbf{b^{U}}$ and $\mathbf{b^V}$ as conditions.
% \begin{equation}
% \resizebox{.91\linewidth}{!}{$
%     \log p(r_{i,j})=\log\sum_{\mathbf{b_i^U},\mathbf{b_j^V \in \{-1,1\}^D}}p(r_{i,j}|\mathbf{b_i^U},\mathbf{b_j^V})p(\mathbf{b_i^U})p(\mathbf{b_j^V}).
%     \label{rating_likelihood}
% $}
% \end{equation}
\begin{equation}
     p(r_{i,j}) = \sum_{\mathbf{b_i^U},\mathbf{b_j^V \in \{-1,1\}^D}}p(r_{i,j}|\mathbf{b_i^U},\mathbf{b_j^V})p(\mathbf{b_i^U})p(\mathbf{b_j^V})
    \label{rating_likelihood}
\end{equation}

However, such maximization is intractable, thus we need to introduce variational inference \cite{Jordan1999}. Variational inference devotes to approximate both posteriors $q_{\phi}(\mathbf{b_{i}^U}|\mathbf{r_{i,*}})$ and $q_{\psi}(\mathbf{b_{j}^V}|\mathbf{r_{*,j}})$ to surrogate the true intractable posterior $p(\mathbf{b_{i}^U}|\mathbf{r_{i,*}})$ and $p(\mathbf{b_{j}^V}|\mathbf{r_{*,j}})$, where $\phi$ and $\psi$ are parameters on neural networks respectively for user and item. Formally, variational inference in this work seeks to simultaneously minimize the KL divergence KL$(q_{\phi}(\mathbf{b_{i}^U}|\mathbf{r_{i,*}})\|p(\mathbf{b_{i}^U}|\mathbf{r_{i,*}})$ and KL$(q_{\psi}(\mathbf{b_{j}^V}|\mathbf{r_{*,j}}\|p(\mathbf{b_{j}^V}|\mathbf{r_{*,j}})$ to achieve the optimal $\phi$ and $\psi$. During the learning process with the above variational inference, an Evidence Lower BOund can be drawn in Eq.(\ref{ELBO}). Therefore, we can use ELBO maximization as a proxy to indirectly maximize the log-likelihood function.
\begin{equation}
\resizebox{.91\linewidth}{!}{$
\begin{aligned}
\log p(r_{i,j}) \geq
\sum_{i,j}\mathbb{E}_{q_{\phi}(\mathbf{b_{i}^U}|\mathbf{r_{i,*}})}\mathbb{E}_{q_{\psi}(\mathbf{b_{j}^V}|\mathbf{r_{*,j}})}[\log\ p(r_{i,j}|\mathbf{b_i^U},\mathbf{b_j^V})]  \\
    -\sum_i\rm{KL}(q_{\phi}(\mathbf{b_i^U}|\mathbf{r_{i,*}})\|p(\mathbf{b_i^U}))-\sum_j\rm{KL}(q_{\psi}(\mathbf{b_j^V}|\mathbf{r_{*,j}})\|p(\mathbf{b_j^V}))
\end{aligned}
$}
\label{ELBO}
\end{equation}
where the first term in ELBO indicates the reconstruction error that measures the likelihood of reconstructing the observed rating data, the two KL terms are regularizers that constrain the form of the two approximate posteriors.

\nosection{Calculation}
In this part, user side and item side are treated symmetrically. Due to the limited space, this part only takes user side as a specific example for illustration. Firstly, we adopt $\bm{E}_U$ to generate user mean and variance as $[\bm{\mu}^U_i,(\bm{\sigma}^U_i)^2] = \bm{E}_U(\bm{r}_{i*})$. Then we adopt the reparametric method to obtain the user real-valued representation as $\bm{z}^U_{i,0} = \bm{\mu}^U_i + \epsilon^U_i \bm{\sigma}^U_i$
%
% \begin{equation}
% \begin{aligned}
% \label{equ:user_rep}
% \bm{z}^U_{i,0} = \bm{\mu}^U_i + \epsilon^U_i \bm{\sigma}^U_i,
% \end{aligned}
% \end{equation}
%\
where $z_{i,0}^U$ is a $D$-dimensional vector. In order to enhance the model generalization, we align the real-valued latent space to the standard normal distribution $\mathcal{N}(0,\bm{I})$ with the KL-divergence constraint as:
\begin{equation}
\small
\begin{aligned}
\label{equ:l_align}
&\min \mathcal{L}_{Align}^U = \min {\rm KL}(\mathcal{N}(\bm{\mu}^U_i,\bm{(\sigma}^U_i)^2)||\mathcal{N}(0,\bm{I})) \\&= \frac{1}{2}\sum_{i=1}^N \left[(\bm{\mu}^U_i)^2 + (\bm{\sigma}^U_i)^2 - \log (\bm{\sigma}^U_i)^2 - 1 \right]
\end{aligned}
\end{equation}

After that we adopt the normalized flow to generate the $\bm{z}^U_{i,T}$ through several layers with probability estimation as:
\begin{equation}
\resizebox{.91\linewidth}{!}{$
\begin{aligned}
\label{equ:probability_estimation}
\log q_{\phi}(\bm{z}^U_{i,T}|_d) = \log q_{\phi}(\bm{z}^U_{i,0}|_d) - \sum_{i=1}^{T-1}\log \left| \det \frac{d\bm{z}^U_{i,t+1}|_d}{d\bm{z}^U_{i,t}|_d}\right|
\end{aligned}
$}
\end{equation}
where $\bm{z}^U_{i,t} + \bm{u}^U_t \chi \left((\bm{w}^U_t)^{\top} \bm{z}^U_{i,t} + \bm{a}^U_t \right) = \bm{z}^U_{i,t+1} \in \mathbb{R}^D$ and $\chi(\cdot)$ denotes the sigmoid activation function.
$\bm{u}^U_t$, $\bm{w}^U_t$ and $\bm{a}^U_t$ are the trainable network parameters at the $t$-th layer.
Here, it is worth noting that the reversible distribution transformation in Eq.(\ref{equ:probability_estimation}) is only deployed on the $d$-th dimension of the entire $D$-dimensional representation.
Finally, we adopt ${\rm sign}(\cdot)$ to achieve the user hash representation as:
\begin{equation}
\begin{aligned}
\label{equ:user_hash_rep}
\bm{b}^U_i = {\rm sign}(\bm{z}^U_{i,T})
\end{aligned}
\end{equation}

Meanwhile we prefer that the latent hash representation conforms to the Bernoulli distribution as $p(\bm{b}_i|_d = 1) = p(\bm{b}_i|_d = -1) = \frac{1}{2}$ on the $d$-th dimension.
However, the Bernoulli distribution is discrete which makes it much more difficult to be optimized.
Therefore, we first innovatively propose the \textit{approximate mixture multivariate normal distribution} represented as:
\begin{equation}
\resizebox{.91\linewidth}{!}{$
\begin{aligned}
\label{equ:distribution}
p(\bm{b}^U_i|_d) \approx p(\bm{z}^U_{i,T}|_d) = \frac{1}{2}\left[ \mathcal{N}(\bm{1},\bm{\Sigma}(b^U_i|_d)) +  \mathcal{N}(\bm{-1},\bm{\Sigma}(b^U_i|_d)) \right]
\end{aligned}
 $}
\end{equation}
where $\bm{\Sigma}(b_i|_d)  = \gamma\bm{I}$ denotes the covariance matrix of corresponding normal distribution.
Here, $\gamma$ is set to 0.015 in our experiments.
%to make the distribution more concentrate on $-1$ and $1$, so as to the distribution is more similar to the original discrete Bernoulli distribution.
%
Therefore, we define the loss function corresponding to the KL term in Eq. (\ref{ELBO}) as:
%
% \begin{equation}
% \resizebox{.91\linewidth}{!}{$
% \begin{aligned}
% \label{equ:kl_term}
% &\mathcal{L}_{\rm KL}^{U}=\sum_i {\rm KL}(q_{\phi}(\mathbf{b_i^U}|\mathbf{r_{i,*}})\|p(\mathbf{b_i^U})) \approx \sum_i \sum_d {\rm KL}(q_{\phi}(\bm{z}^U_{i,t}[d])\|p(\bm{z}^U_{i,T}[d])) \\& = \sum_i \sum_d q_{\phi}(\bm{z}^U_{i,t}[d]) \log \frac{q_{\phi}(\bm{z}^U_{i,t}[d])}{p(\bm{z}^U_{i,T}[d])}.
% \end{aligned}
%  $}
% \end{equation}

\begin{equation}
\begin{aligned}
\label{equ:kl_term}
&\mathcal{L}_{\rm KL}^{U}=\sum_i {\rm KL}(q_{\phi}(\mathbf{b_i^U}|\mathbf{r_{i,*}})\|p(\mathbf{b_i^U})) \\ & \approx \sum_i \sum_d q_{\phi}(\bm{z}^U_{i,t}|_d) \left[\log q_{\phi}(\bm{z}^U_{i,t}|_d) - \log p(\bm{z}^U_{i,T}|_d)\right]
\end{aligned}
\end{equation}
Here, since $p(\bm{b}^U_i)$ is approximately surrogated by an \textit{approximate mixture multivariate normal distribution} in Eq.(\ref{equ:distribution}), the $\rm KL$ term is also be approximated in Eq.(\ref{equ:kl_term}).

Next, we infer the reconstruction loss corresponding to the likelihood $\mathcal{L}_{recon} = -p(r_{i,j}|\mathbf{b_i^U}, \mathbf{b_j^V})$ in Eq.(\ref{ELBO}). In this paper, we assume that the observed rating data obey the Poission distribution, such that:
% \begin{equation}
% \resizebox{\linewidth}{!}{$
%   \mathcal{L}_{recon}=-p(r_{i,j}|\mathbf{b_i^U},\mathbf{b_j^V})=-\frac{1}{r_{i,j}!}\exp\left\{r_{i,j}\log\left(\frac{\mathbf{(b_i^U)}^{T}\mathbf{b_j^V}+K}{2K}\right)-\frac{\mathbf{(b_i^U)}^{T}\mathbf{b_j^V}+K}{2K}\right\}.
%  $}
% \label{hpois}
% \end{equation}

\begin{equation}
\begin{aligned}
  \mathcal{L}_{recon}&=-\frac{1}{r_{i,j}!}\exp\left(r_{i,j}\log\left( s_B(b_i^U,b_j^V)  \right) - s_B(b_i^U,b_j^V)\right) .
\label{hpois}
\end{aligned}
\end{equation}
where $ s_B(b_i^U,b_j^V) = \frac{\mathbf{(b_i^U)}^{T}\mathbf{b_j^V}+K}{2K}$
Considering the equivalence of inner product and Hamming distance, hash representations can be optimized directly by optimizing Eq.(\ref{hpois}).

Finally, fusing both user side and item side information, the loss function for the optimal hash representations can be defined as:
\begin{equation}
 \mathcal{L}_{rating}= \mathcal{L}_{recon}+\mathcal{L}_{Align}^U+\mathcal{L}_{KL}^{U}+\mathcal{L}_{Align}^V+\mathcal{L}_{KL}^{V}
 \label{rating_loss}
\end{equation}

It is worth noting that, in Eq.(\ref{equ:user_hash_rep}), $\bm{b}^U_i$ ($\bm{b}^V_j$) is achieved by ${\rm sign}(\cdot)$, a non-smooth function that makes the gradient of all inputs to zero during backward propagation. Thus, we adopt an identify function $f(\cdot)$ to surrogate $\rm sign(\cdot)$, such that $f(\cdot)$ achieves unit gradient in the backward pass.

\subsection{Cluster Consistency Preserving Component}
\label{3.2}
 To further enhance the quality of both real-valued representations and hash representations, we push them to reach structural consensus. In other words, neighbors generated by real-valued representations and hash representations are expected to be same, such that semantic cluster structure \cite{liu2021leveraging} can be preserved. \\
\nosection{Neighbor Aggregation} In this part, since user side and item side can work in the same way, thus, for concise illustration, we employ user side as a specific example for illustration. Firstly, we aggregate neighbors into different clusters respectively with real-valued representations and hash representations. Due to the characteristics of data independence and time-efficiency, Locality Sensitivity Hashing (LSH) has been proven a powerful approach for approximate nearest neighbor (ANN) search \cite{Qi2021}. Thus, we resort to LSH for neighbor aggregation. Concretely, since the intrinsic continuity of real-valued representations and the discreteness of hash representations, the hash functions of the two representations are respectively for Euclidean distance and Hamming distance. Formally, for the real-valued side, we recruit hash functions as:
\begin{equation}
    h_i(\bm{z^U_{i,0}})=\lfloor\frac{\bm{a}\cdot \bm{z^U_{i,0}}+c}{w}\rfloor
    \label{real_lsh}
\end{equation}
where $\bm{a}\sim \mathcal{N}(0,I)$ is a $K$-dimentional random vector, $c\sim \mathcal{U}(0,w)$ is a random real value, and $w\in \mathbb{R}^{+}$ is a hyper-parameter. Since LSH is a probability-based approach, we perform the above hash process $L$ rounds to ensure credibility. Then, the final hash value can be calculated by:
\begin{equation}
    H_i(\bm{z^U_{i,0}})=\sum_{l=1}^{L}B^{l}h_i^{(l)}(\bm{z^U_{i,0}})
\label{final_hash1}
\end{equation}
where $B$ is a constant, and $h_i^{(l)}$ is the hash function employed in round $l$ with independently sampled random variables $\bm{a}$ and $c$. Then, users with the same hash values will be projected into an identical bucket. We regard neighbors in the same bucket as a cluster. From the obtained cluster structure, we can obtain the similarity matrix $A\in \{1,0\}^{N_U\times N_U}$ corresponding to real-valued latent representations as:
\begin{equation}
   A_{i_1,i_2}=\left\{
\begin{aligned}
1,\ \ \ \  H_{i_1}(\bm{z^U_{i_1,0}})= H_{i_2}(\bm{z^U_{i_2,0}}) \\
0,\ \ \ \  H_{i_1}(\bm{z^U_{i_1,0}})\neq H_{i_2}(\bm{z^U_{i_2,0}})\\
\end{aligned}
\right.
\label{similar_matrix1}
\end{equation}
where 1 indicates ``similar", 0 indicates not. Likewise, for hash representations, we perform the hash functions for Hamming distance $L$ rounds and calculate the final hash values by Eq.(\ref{final_hash2}).
\begin{equation}
 \resizebox{.8\linewidth}{!}{$
    h_i^{'(l)}(\bm{b_i^U})=\bm{b_i}^{(d)}, \quad H'_i(\bm{b_i^U})=\sum_{l=1}^L 2^{L-1}h_i^{'(l)}
    \label{final_hash2}
     $}
\end{equation}
where $h_i^{'(l)}$ denotes hash functions, $H'_i(\bm{b_i^U})$ denotes final hash values, $\bm{b_i}^{(d)}$ indicates the binary value of the $d$-th dimension in $D$-dimensional hash representation $\bm{b_i}$.
Users (items) with the same hash values are projected into an identical bucket and regarded as a cluster. With the obtained cluster structure, the similarity matrix $A'\in \{1,0\}^{N_U\times N_U}$ corresponding to hash representations can also be obtained.

\nosection{Consistency Learning} Till now, joining both user side and item side,  the optimization objective of cluster consistency can be defined by:
\begin{equation}
    \resizebox{.91\linewidth}{!}{$
\begin{aligned}
    \mathcal{L}_{cons}=\sum_{i_1=1}^{N_U}\sum_{i_2=1}^{N_U} \|\bm{z_{i_1}^U}-\bm{z_{i_2}^U}\|_2A_{i_1,i_2}^{'}
    + \sum_{i_1=1}^{N_U}\sum_{i_2=1}^{N_U}\bm{(b_{i_1}^U)^T}\bm{b_{i_2}^U}A_{i_1,i_2}\\
    +\sum_{j_1=1}^{N_V}\sum_{j_2=1}^{N_V} \|\bm{z_{j_1}^V}-\bm{z_{j_2}^V}\|_2A_{j_1,j_2}^{'}
    + \sum_{j_1=1}^{N_V}\sum_{j_2=1}^{N_V}\bm{(b_{j_1}^V)^T}\bm{b_{j_2}^V}A_{j_1,j_2}
\end{aligned}
    $}
\label{cons_loss}
\end{equation}
where, for the user side, the first term indicates that similar users' hash representations require smaller Euclidean distances for users' real-valued representations, while the second term indicates that similar users' real-valued representations require smaller Hamming distances for users' hash representations, which is likewise for the item side.

\subsection{Combined Loss Function}
Next, we put the two components together and get a fused loss function of HCFRec defined as:
\begin{equation}
    \mathcal{L}=\mathcal{L}_{rating}+\lambda\mathcal{L}_{cons}
\label{lossforall}
\end{equation}
where $\lambda$ is a hyper-parameter that balances the loss of two components, $\mathcal{L}_{rating}$ is the loss of the first component obtained in Eq.(\ref{rating_loss}), and $\mathcal{L}_{cons}$ is the loss of the second component obtained in Eq.(\ref{cons_loss}).
\section{Experiments and Evaluation}
We conduct extensive experiments to answer the following questions: \textbf{Q1:} How does HCFRec compare with the state-of-art approaches in terms of recommendation accuracy (see subsection \ref{5.3})? \textbf{Q2:} How does Hamming distance outperform real-valued inner product in terms of computational efficiency and storage costs (see subsection \ref{5.4})? \textbf{Q3:} How does HCFRec perform on parameter sensitivity (see subsection \ref{5.5})?
%\textbf{Q4:} How does the two key modules contribute to CH-VAE's performance? \textcolor[rgb]{1,0,0}{ablation study?}\\
\subsection{Data Preparation}
We adopt two well-known datasets, MovieLens\footnote{https://grouplens.org/datasets/movielens} and Amazon\footnote{http://jmcauley.ucsd.edu/data/amazon/} for experimental evaluations: (1) \textbf{MovieLens} collects user ratings for movies, ranging from 1 (worst) to 5 (best). We evaluate recommendation performance in terms of different data scale, i.e., \emph{ML-100K}, \emph{ML-1M}, \emph{ML-10M}. (2) \textbf{Amazon} \cite{He2016Ups} covers ratings (with the range of 1 to 5) for up to 24 product categories. We evaluate recommendation performance on its 3 product categories, i.e., \emph{Clothing, Shoes and Jewelry}, \emph{Office Products}, and \emph{Toys and Games}.

We preprocess the data following \cite{Hansen2020Content} and \cite{Lian2017DiscreteCM} to filter users and items with less than 20 ratings. Moreover, only the last rating is reserved if a user has rated an item multiple times. We sort the ratings in ascending order according to the feedback time, and divide all datasets into training set, validation set, and test set according to 5:2:3. We summarize the dataset information with Table \ref{tab:datasets}.
% We specify the information of the filtered and experimental datasets in Table \ref{tab:datasets}.
\renewcommand\arraystretch{0.8}
\begin{table}[t]
\setlength{\abovecaptionskip}{0.cm}
\centering
\resizebox{\linewidth}{!}{
\begin{tabular}{llllll}
\toprule
  \multicolumn{2}{c}{Datasets}   &  \#users&  \#itmes& \#ratings  &\#sparsity\\
\midrule
             & ML-100K\hspace{1mm} &911     & 927     & 47,056     & 94.428\% \\
MovieLens    & ML-1M              &5,954    & 3,011    & 496,573    & 97.230\% \\
             & ML-10M             &67,976   & 8,882    & 4,972,679   & 99.176\% \\
\midrule
             & Clothing           &6,671    & 20,755   & 39,819     & 99.971\% \\
Amazon       & Office             &838     & 2,894    & 11,385     & 99.531\% \\
             & Toys               &2,634    & 10,059   & 26,293     & 99.901\% \\
\bottomrule
\end{tabular}}
\caption{Datasets descriptions.}
\label{tab:datasets}
\end{table}

% \begin{table}[htbp]
% \caption{Mean forecast errors of the three methods} \label{speedup}\tabcolsep
% 0pt \vspace*{-10pt}
% \par
% \begin{center}
% \def\temptablewidth{0.5\textwidth}
% {\rule{\temptablewidth}{1pt}}
% \begin{tabular*}{\temptablewidth}{@{\extracolsep{\fill}}l l |llll}
% \hline
%   Dataset    &           &  \#users&  \#itmes& \#ratings  &\#sparsity\\
% \hline
%              & ML-100K\hspace{1mm} &911     & 927     & 47,056     & 94.428\% \\
%  MovieLens   & ML-1M              &5,954    & 3,011    & 496,573    & 97.230\% \\
%              & ML-10M             &67,976   & 8,882    & 4,972,679   & 99.176\% \\
% \hline
%              & Clothing           &6,671    & 20,755   & 39,819     & 99.971\% \\
% Amazon       & Office             &838     & 2,894    & 11,385     & 99.531\% \\
%              & Toys               &2,634    & 10,059   & 26,293     & 99.901\% \\
% \hline
% \label{dataset}
% \end{tabular*}%
% \end{center}
% \end{table}

%\vspace{-0.3cm}
\subsection{Experimental Settings}
\nosection{Comparison Methods}
Five baselines classified into two groups and two versions of our proposed approaches, i.e.,  HCFRec/no.C and HCFRec, are compared in our experiments, where HCFRec/no.C is the version of HCFRec without cluster consistency preserving component.
% We employ two classical real-valued CF approaches as benchmark and compare our H-VAE as well as CH-VAE with two state-of-the-art Hash-CF approaches.

(1) \emph{Classical real-valued CF approaches}:
\textbf{BPR}: A classical recommendation framework maximizing posterior estimator to create a personalized ranking list for a group of items \cite{Rendle2009}.
\textbf{BiVAE}: A state-of-the-art bilateral VAE that learns user real-valued representations and item real-valued representations for CF \cite{Truong2021}\footnote{https://github.com/PreferredAI/bi-vae}.

(2) \emph{Hash-CF approaches}:
\textbf{DCF}: An explicit feedback-based Hash-CF method that learns balanced and uncorrelated hash codes for recommendation task \cite{Zhang2016DCF}\footnote{https://github.com/hanwangzhang/Discrete-Collaborative-Filtering}.
\textbf{DPR}: A Hash-CF method tha learns hash codes based on personalized ranking objective instead of rating prediction objective in DCF \cite{Zhang2017DiscretePR}\footnote{https://github.com/yixianqianzy/dpr}.
\textbf{BiVAEB}: A version directly binarizing the real-valued representations in BiVAE for recommendation task, which is a baseline indicating the quantization loss from real values to binary representations.

\nosection{Evaluation Metrics}
We focus on the item positions in recommendation lists to evaluate the accuracy of the above competitive methods. In concrete, we adopt the following two metrics: mean Average Precision (mAP) and normalized Discounted Cumulatiive Gain (nDCG).
\renewcommand\arraystretch{0.8}
\begin{table*}[!t]
\scriptsize
% \footnotesize
\centering
%\vspace{-0.2cm}

\begin{tabular}{M{1.1cm}M{0.4cm}M{0.4cm}M{0.4cm}M{0.4cm}M{0.4cm}M{0.4cm}M{0.4cm}M{0.4cm}M{0.4cm}M{0.4cm}M{0.4cm}M{0.4cm}M{0.4cm}M{0.4cm}M{0.4cm}M{0.4cm}M{0.4cm}M{0.4cm}}
\toprule
\multirow{1}{*}{}
& \multicolumn{6}{c}{MovieLens-100K}
& \multicolumn{6}{c}{MovieLens-1M}
& \multicolumn{6}{c}{MovieLens-10M}\\

\cmidrule(lr){2-4} \cmidrule(lr){5-7} \cmidrule(lr){8-10}
\cmidrule(lr){11-13} \cmidrule(lr){14-16} \cmidrule(lr){17-19}
&\multicolumn{3}{c}{16dim}
&\multicolumn{3}{c}{64dim}
&\multicolumn{3}{c}{16dim}
&\multicolumn{3}{c}{64dim}
&\multicolumn{3}{c}{16dim}
&\multicolumn{3}{c}{64dim}
\\

\cmidrule(lr){2-4} \cmidrule(lr){5-7} \cmidrule(lr){8-10}
\cmidrule(lr){11-13} \cmidrule(lr){14-16} \cmidrule(lr){17-19}
nDCG
&@2 &@6 &@10
&@2 &@6 &@10
&@2 &@6 &@10
&@2 &@6 &@10
&@2 &@6 &@10
&@2 &@6 &@10
%BPR&  @2   &   @6  &  @10  &   @2  &    @6 &   @10 &   @2  &  @6   &  @10    &   @2  &    @6  &   @10  &  @2   &    @6 &    @10  &   @2  &    @6  &  @10 \\
\\\midrule
BPR     & .4840 & .4644 & .4498 & .4815 & .4468 & .4231 & .5194 & .4944 & .4765   & .4607 &  .4438 & .4290  & .5607 & .5241 & .5002   & .5619 &  .5285 & .5069\\
BiVAE   & .6599 & .6031 & .5692 & .6415 & .5919 & .5580 & .5953 & .5611 & .5362   & .6008 &  .5609 & .5364  & .5665 & .5301 & .5063   & .5542 &  .5118 & .4857\\
\hline
DCF     & .3031 & .2968 & .2854 & .3279 & .3043 & .2966 & .2436 & .2393 & .2342   & .3743 &  .3559 & .3487  & .1297 & .1364 & .1183   & .1836 &  .1792 & .1723\\
DPR     & .3577 & .3423 & .3389 & .3821 & .3702 & .3572 & .2607 & .2526 & .2460   & .3922 &  .3687 & .3614  & .1305 & .1402 & .1394   & .2157 &  .2248 & .2186\\
BiVAEB  & .2254 & .2157 & .2059 & .1656 & .1557 & .1484 & .1821 & .1713 & .1638   & .0958 &  .0907 & .0870  & .1179 & .1083 & .1019   & .0986 &  .0908 & .0861\\
\hline
HCFRec/no.C  & .4039 & .3945 & .3815 & .4429 & .4072 & .3849 & .2708 & .2547 & .2503   & .4169 &  .3892 & .3725  & .1334 & .1448 & .1575    & .2371 &  .2458 & .2447\\
HCFRec& \textbf{.4599} & \textbf{.4155} & \textbf{.3997} & \textbf{.4709} & \textbf{.4410} & \textbf{.4158} & \textbf{.3638} & \textbf{.3432} & \textbf{.3293}   & \textbf{.4425} &  \textbf{.4099} & \textbf{.3895}  & \textbf{.1814} & \textbf{.1872} & \textbf{.1943}    & \textbf{.2775} &  \textbf{.2698} & \textbf{.2539}\\
%\bottomrule
\toprule
\multirow{1}{*}{}
& \multicolumn{6}{c}{Amazon-Clothing}
& \multicolumn{6}{c}{Amazon-Office}
& \multicolumn{6}{c}{Amazon-Toys}\\
\cmidrule(lr){2-4} \cmidrule(lr){5-7} \cmidrule(lr){8-10}
\cmidrule(lr){11-13} \cmidrule(lr){14-16} \cmidrule(lr){17-19}
&\multicolumn{3}{c}{16dim}
&\multicolumn{3}{c}{64dim}
&\multicolumn{3}{c}{16dim}
&\multicolumn{3}{c}{64dim}
&\multicolumn{3}{c}{16dim}
&\multicolumn{3}{c}{64dim}
\\
\cmidrule(lr){2-4} \cmidrule(lr){5-7} \cmidrule(lr){8-10}
\cmidrule(lr){11-13} \cmidrule(lr){14-16} \cmidrule(lr){17-19}
nDCG
&@2 &@6 &@10
&@2 &@6 &@10
&@2 &@6 &@10
&@2 &@6 &@10
&@2 &@6 &@10
&@2 &@6 &@10
%BPR&  @2   &   @6  &  @10  &   @2  &    @6 &   @10 &   @2  &  @6   &  @10    &   @2  &    @6  &   @10  &  @2   &    @6 &    @10  &   @2  &    @6  &  @10 \\
\\\midrule
BPR     & .0047 & .0041 & .0044 & .0056 & .0054 & .0056 & .0512 & .0504 & .0485   & .0818 &  .0761 & .0738  & .0154 & .0143 & .0136   & .0275 &  .0260 & .0256\\
BiVAE   & .0073 & .0067 & .0071 & .0089 & .0076 & .0080 & .0973 & .0867 & .0816   & .0803 &  .0792 & .0729  & .0390 & .0352 & .0329   & .0364 &  .0326 & .0312\\
\hline
DCF     & .0015 & .0019 & .0018 & .0019 & .0021 & .0019 & .0256 & .0227 & .0210   & .0496 &  .0421 & .0398  & .0095 & .0089 & .0087   & .0187 &  .0180 & .0179\\
DPR     & .0016 & .0021 & .0022 & .0026 & .0025 & .0025 & .0267 & .0240 & .0239   & .0573 &  .0564 & .0552  & .0103 & .0096 & .0093   & .0213 &  .0209 & .0199\\
BiVAEB  & .0006 & .0005 & .0006 & .0002 & .0004 & .0005 & .0132 & .0143 & .0130   & .0220 &  .0207 & .0194  & .0075 & .0063 & .0070   & .0047 &  .0045 & .0047\\
\hline
HCFRec/no.C  & .0022 & .0026 & .0026 & .0030 & .0027 & .0030 & .0270 & .0332 & .0309   & .0693 &  .0620 & .0591  & .0128 & .0108 & .0098   & .0233 &  .0213 & .0206\\
HCFRec& \textbf{.0037} & \textbf{.0029} & \textbf{.0034} & \textbf{.0041} & \textbf{.0035} & \textbf{.0037} & \textbf{.0478} & \textbf{.0474} & \textbf{.0478}   & \textbf{.0779} &  \textbf{.0748} & \textbf{.0723}  & \textbf{.0134} & \textbf{.0131} & \textbf{.0135}   & \textbf{.0242} &  \textbf{.0224} & \textbf{.0221}\\
\bottomrule
\end{tabular}
\setlength{\abovecaptionskip}{-0.cm}
\caption{Experimental results on datasets.}
\label{tab:result1}
\end{table*}
\renewcommand\arraystretch{0.8}
\begin{table*}[!t]
\scriptsize
% \footnotesize
\centering
\begin{tabular}{M{1.1cm}M{0.7cm}M{0.7cm}M{0.7cm}M{0.7cm}M{0.7cm}M{0.7cm}M{0.7cm}M{0.7cm}M{0.7cm}M{0.7cm}M{0.7cm}M{0.7cm}}
\toprule
\multirow{1}{*}{}
& \multicolumn{2}{c}{MovieLens-100K}
& \multicolumn{2}{c}{MovieLens-1M}
& \multicolumn{2}{c}{MovieLens-10M}
& \multicolumn{2}{c}{Amazon-Clothing}
& \multicolumn{2}{c}{Amazon-Office}
& \multicolumn{2}{c}{Amazon-Toys}\\
\cmidrule(lr){2-3} \cmidrule(lr){4-5} \cmidrule(lr){6-7}
\cmidrule(lr){8-9} \cmidrule(lr){10-11} \cmidrule(lr){12-13}
mAP@10&16dim &64dim
&16dim &64dim
&16dim &64dim
&16dim &64dim
&16dim &64dim
&16dim &64dim
\\\midrule
BPR     & .0801 & .0772 & .0556 & .0520 & .0865 & .0917 & .0016 & .0021 & .0112   & .0189 &  .0032 & .0071 \\
BiVAE   & .1159 & .1136 & .0699 & .0713 & .0909 & .0832 & .0025 & .0030 & .0213   & .0182 &  .0093 & .0096  \\
\hline
DCF     & .0427 & .0431 & .0135 & .0289 & .0108 & .0193 & .0005 & .0007 & .0054   & .0125 &  .0013 & .0049  \\
DPR     & .0496 & .0501 & .0143 & .0316 & .0126 & .0208 & .0006 & .0008 & .0059   & .0133 &  .0015 & .0053 \\
BiVAEB  & .0227 & .0139 & .0100 & .0055 & .0060 & .0059 & .0002 & .0001 & .0026   & .0043 &  .0011 & .0013  \\
\hline
HCFRec/no.C & .0581 & .0585 & .0176 & .0369 & .0131 & .0239 & .0009 & .0010 & .0062   & .0148 &  .0024 & .0061  \\
HCFRec& \textbf{.0642} & \textbf{.0670} & \textbf{.0283} & \textbf{.0394} & \textbf{.0184} & \textbf{.0293} & \textbf{.0013} & \textbf{.0014} & \textbf{.0106}   & \textbf{.0175} &  \textbf{.0026} & \textbf{.0064} \\
\bottomrule
\end{tabular}
\setlength{\abovecaptionskip}{-0cm}
\caption{Experimental results on datasets.}
\label{tab:result2}
\end{table*}

\nosection{Parameter Setting}
We conduct our experiments on an NVIDIA RTX 3090 GPU by PyTorch. We adopt Adam optimizer with learning rate 0.015 for training. Moreover, the mini-batch SGD with the fixed batch size 128 is employed for optimization. As the introduced hyper-parameters in our model, we set $\gamma$ in Eq. (\ref{equ:distribution}) as 0.015 and $\lambda=0.3$ in Eq.(\ref{lossforall}). Following \cite{Zheng2020EndtoEnd}, we set $\omega=8$ in Eq.(\ref{real_lsh}) and $B=4$ in Eq(\ref{final_hash1}). Moreover, we set $L=1$ in both Eq (\ref{final_hash1}) and Eq (\ref{final_hash2}), because multiple rounds of training have replaced $L$ to make the obtained hash values confident.
%\vspace{-0.2cm}
\subsection{Performance Comparison (for Q1)}
\label{5.3}
We report the recommendation accuracy in terms of nDCG and mAP respectively in Table \ref{tab:result1} and Table \ref{tab:result2}.
 Moreover, we evaluate the performance on all methods with both 16-dimensional and 64-bit representations. To illustrate the performance intuitively, we boldly mark the best performance among all Hash-CF methods in each column.

As Table \ref{tab:result1} and Table \ref{tab:result2} show, HCFRec achieves superior performance over other Hash-CF baselines with improvements of at least 7.18\% in terms of nDCG and at leat 20.75\% in terms of mAP. Moreover, on Amazon-Clothing and Amazon-Toys datasets (two extremely sparse settings), HCFRec also achieves higher robustness of recommendations over other Hash-CF baselines, which demonstrates a better generalization ability of VAE framework. Furthermore, considering the performance gap between HCFRec/no.C and HCFRec, we find that semantic information takes effect for more informative hash codes.

In addition to Hash-CF approaches, the state-of-the-art real-valued CF (BPR and BiVAE) are also employed for experimental evaluation. As expected, the two real-valued CF methods perform better on all datasets since they leverage more informative representations than Hash-CF methods. However, we can observe that the performance gap becomes smaller when dimension increases. This is because even in low-dimensional scenarios, real-valued CF can still collect rich information for recommendations. In this situation, higher dimensions often bring a limited performance improvement. As a contrast, more dimensions of hash code often result in a higher ability of information representation for better recommendation performances.

\subsection{Efficiency and Storage (for Q2)}
\label{5.4}
\begin{figure}[htbp]
%\vspace{-0.7cm}                         %调整图片与上文的垂直距离
\setlength{\abovecaptionskip}{-0.1cm}      %调整图片标题与图距离
\setlength{\belowcaptionskip}{-0.4cm}   %调整图片标题与下文距离
\subfigcapskip=-3pt
\centering
\subfigure[computational efficiency]{
\includegraphics[width=1.6in]{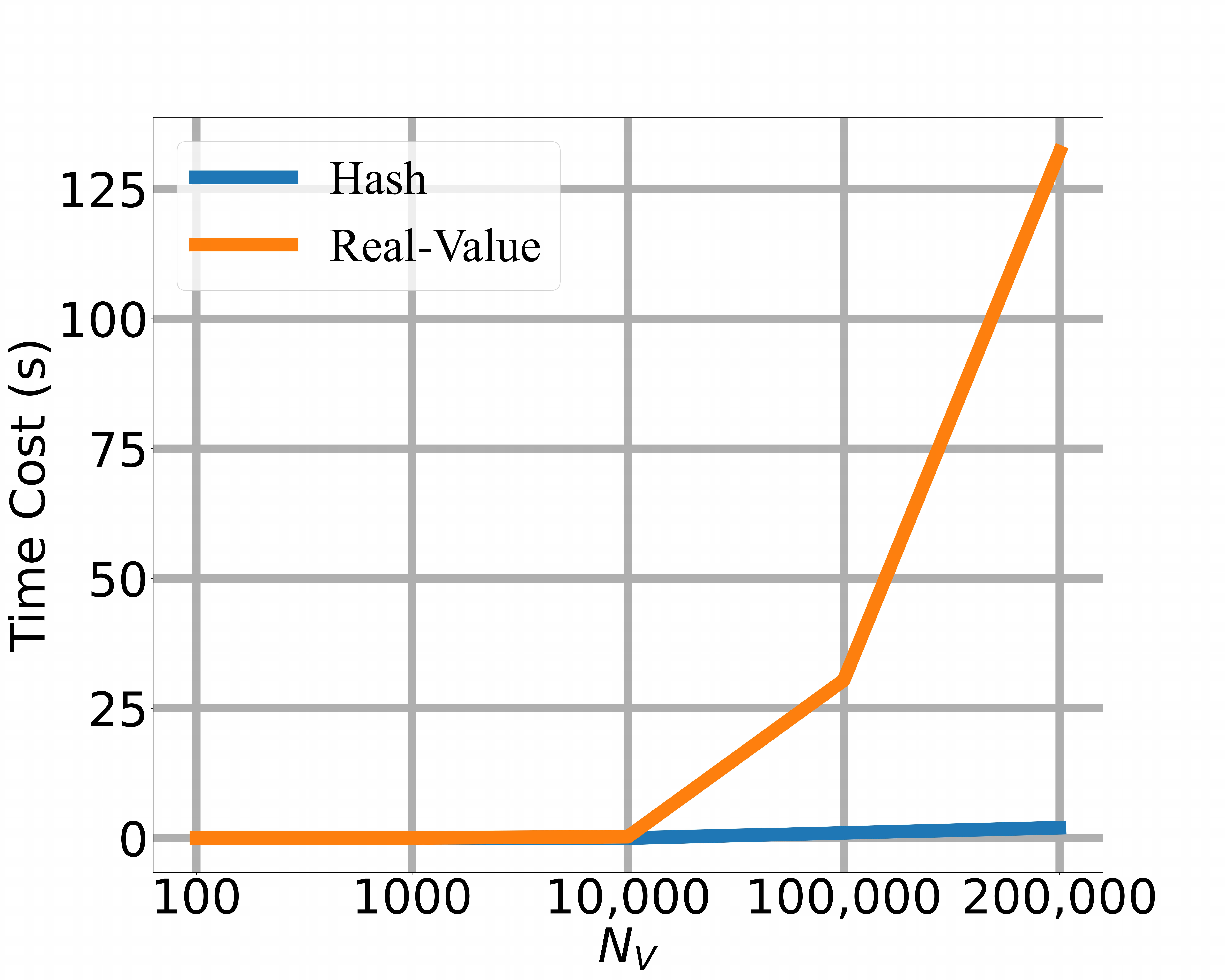}}\label{fig:8a}
\subfigure[storage cost]{
\includegraphics[width=1.6in]{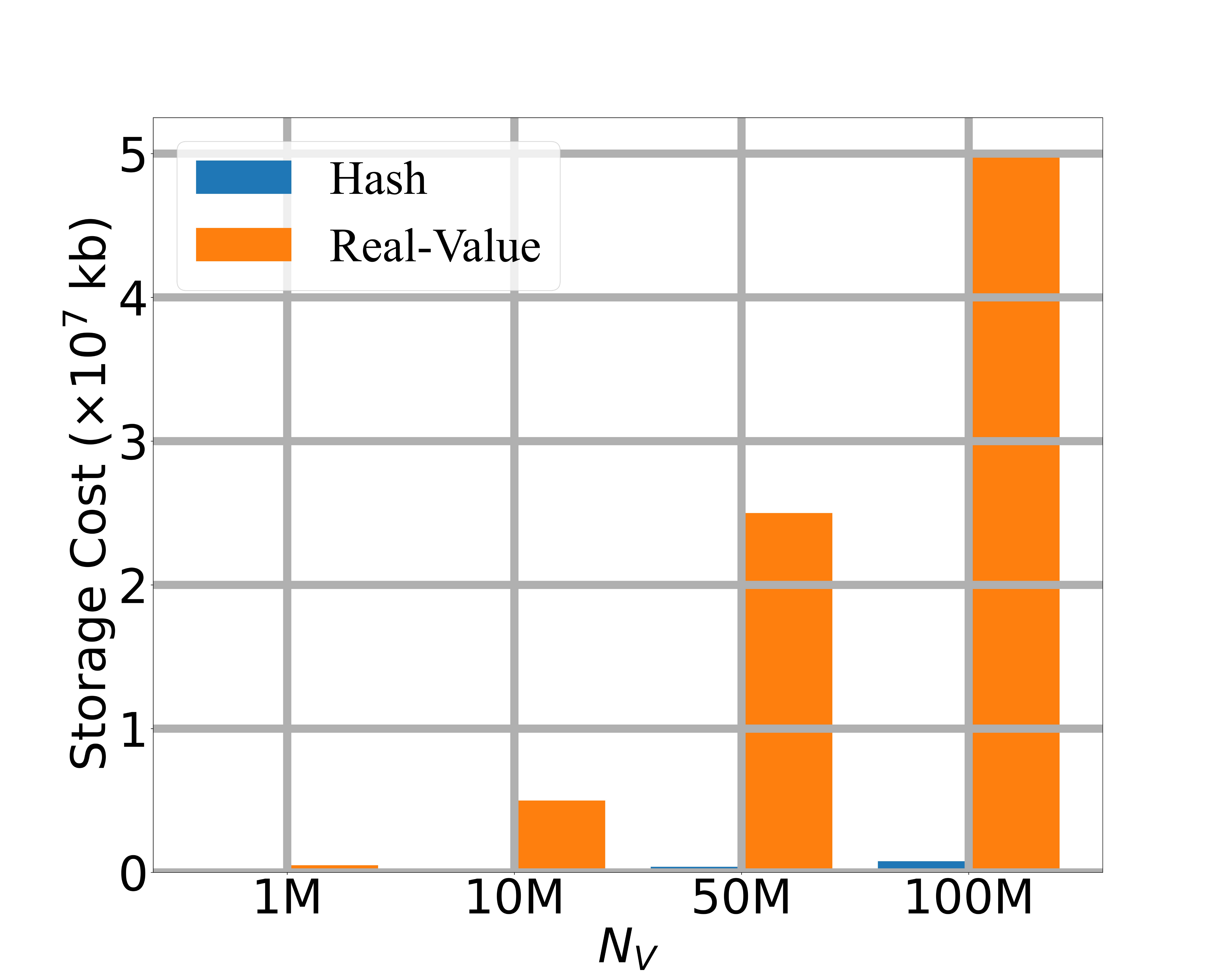}}
\label{fig:8b}
\caption{Comparisons of real-valued representations and hash representations in terms of efficiency and storage.}
\label{efficiency}       % Give a unique label
\end{figure}
Recall that one of the major motivations of our proposal is to recruit hash representation for more efficient computation and lighter storage. In this subsection, we respectively investigate computational efficiency and storage cost as follows.\\
\nosection{Computational Efficiency} We fix the user size to 100,000 and change the item size $N_V$ from 100 to 200,000. The 64-dimensional representations of real-value and hash codes are randomly generated. The evaluation results are illustrated in Figure \ref{efficiency}(a): computational time of real-valued representations grows exponentially as item scale increases; as a contrast, the time cost of hash representations achieves a speedup factor of 40-50. This observation demonstrates the efficiency benefit of Hash-CF for large-scale recommendation.\\
\nosection{Storage Costs} We compare the storage costs in Figure \ref{efficiency}(b) with item sizes from 1 million to 1 billion. As expected, storing real-valued representations consumes more space than hash codes. This is because one dimension of real-valued representation often needs 64 bits, while one dimension of hash code needs only 1 bit. This fact highlights the space thrift of Hash-CF for large-scale recommendation.

\subsection{Parameter Sensitivity  (for Q3)}
\label{5.5}
\begin{figure}[t]
%\vspace{-0.2cm}                         %调整图片与上文的垂直距离
\setlength{\abovecaptionskip}{-0.1cm}      %调整图片标题与图距离
\setlength{\belowcaptionskip}{-0.5cm}   %调整图片标题与下文距离
\subfigcapskip=-3pt %设置子图与子标题之间的距离
\centering
\subfigure[sensitivity of $\gamma$]{
\includegraphics[width=1.6in]{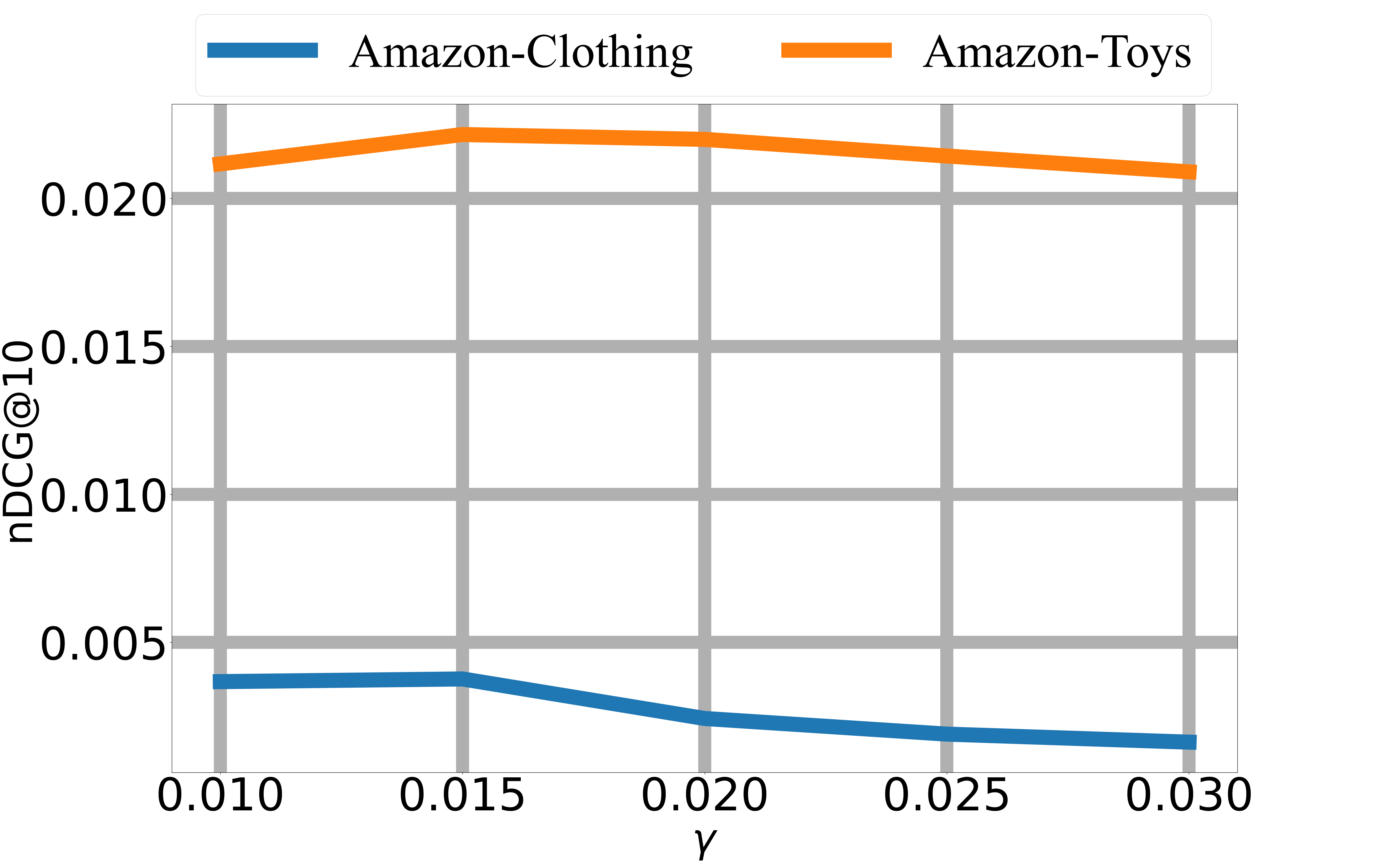}}\label{fig:9a}
\subfigure[sensitivity of $\lambda$]{
\includegraphics[ width=1.6in]{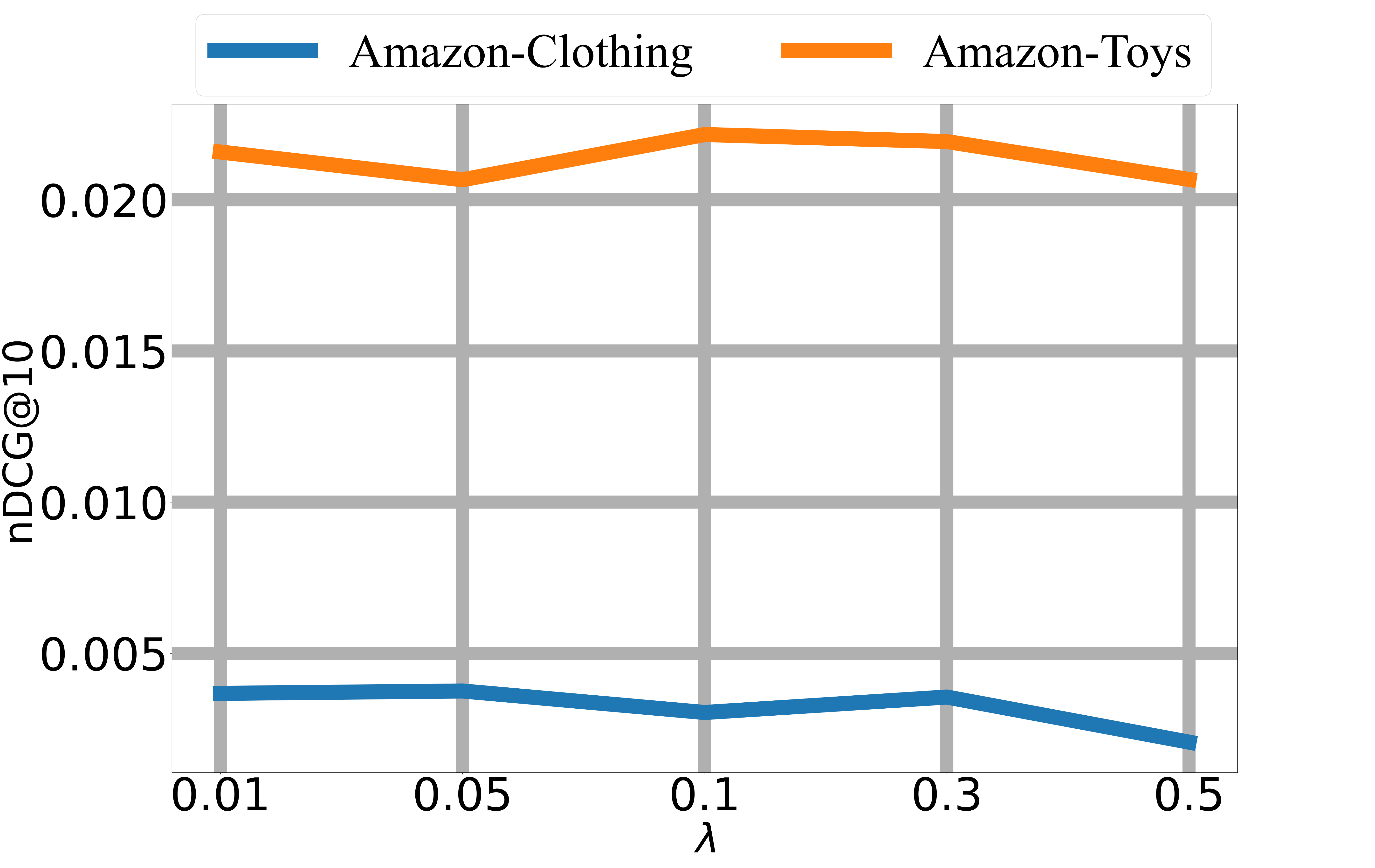}}
\label{fig:9b}
\caption{nDCG@10 w.r.t. different $\gamma$ and $\lambda$.}
\label{para}       % Give a unique label
\end{figure}

We investigate the recommendation performances influenced  by hyper-parameters $\gamma$ and $\lambda$ in Figure \ref{para}, during which the 64-dimensional hash representations are adopted and the top-10 items are returned for evaluation nDCG performance on the two sparsest datasets, i.e., Amazon-Clothing and Amazon-Toys. With fixed $\lambda=0.3$, we vary $\gamma$ from 0.010 to 0.030. Figure \ref{para}(a) shows that our model achieves the best performance when $\gamma=0.015$. This is because a continuous distribution extremely close to a discrete distribution (with $\gamma$ smaller than 0.015) is too complex to learn. As a contrast, the much smoother distribution (with $\gamma$ larger than 0.015) leads to more quantization errors in the latter quantization step, so as to dampen the recommendation performance. Furthermore, we fix $\gamma=0.015$ and evaluate nDCG@10 with $\lambda$ verying from 0.01 to 0.5. Figure \ref{para}(b) shows that the performances are not sensitive to $\lambda$. Therefore, we set $\gamma=0.015$ and $\lambda=0.3$.
\section{Conclusion}
In the big data environment, Hash-CF has been proven a promising technique to accelerate recommendation efficiency by learning an optimal hash representation. However, traditional Hash-CF often falls short in the optimization on discrete hash representations and the preservation of semantic information. To tackle these issues, we introduce normalized flow to learn the optimal hash code and deploy a cluster consistency preserving mechanism to preserve the semantic structure in representations.  Extensive experiments conducted on six real-world datasets reveal the superiority of our proposal in terms of accuracy and efficiency.

\section*{Acknowledgments}

This work was supported in part by the National Key R\&D Program of China (No.2018YFB1403001), the  National Natural Science Foundation of China (No.62172362 and No.72192823) and Leading Expert of ``Ten Thousands Talent Program'' of Zhejiang Province (No.2021R52001).

\bibliographystyle{named}
\bibliography{ijcai22}

\begin{thebibliography}{}

\bibitem[\protect\citeauthoryear{Chen \bgroup \em et al.\egroup
  }{2018}]{Chen2018Distributed}
Chaochao Chen, Ziqi Liu, Peilin Zhao, Longfei Li, Jun Zhou, and Xiaolong Li.
\newblock Distributed collaborative hashing and its applications in ant
  financial.
\newblock In {\em ACM SIGKDD}, 2018.

\bibitem[\protect\citeauthoryear{Chen \bgroup \em et al.\egroup
  }{2022}]{chen2022differential}
Chaochao Chen, Huiwen Wu, Jiajie Su, Lingjuan Lyu, Xiaolin Zheng, and Li~Wang.
\newblock Differential private knowledge transfer for privacy-preserving
  cross-domain recommendation.
\newblock {\em arXiv preprint arXiv:2202.04893}, 2022.

\bibitem[\protect\citeauthoryear{Hansen \bgroup \em et al.\egroup
  }{2020}]{Hansen2020Content}
Casper Hansen, Christian Hansen, Jakob~Grue Simonsen, Stephen Alstrup, and
  Christina Lioma.
\newblock Content-aware neural hashing for cold-start recommendation.
\newblock In {\em ACM SIGIR}, 2020.

\bibitem[\protect\citeauthoryear{He and McAuley}{2016}]{He2016Ups}
Ruining He and Julian McAuley.
\newblock Ups and downs: Modeling the visual evolution of fashion trends with
  one-class collaborative filtering.
\newblock In {\em WWW}, 2016.

\bibitem[\protect\citeauthoryear{Jordan \bgroup \em et al.\egroup
  }{1999}]{Jordan1999}
Michael~I Jordan, Zoubin Ghahramani, Tommi~S Jaakkola, and Lawrence~K Saul.
\newblock An introduction to variational methods for graphical models.
\newblock {\em Machine learning}, 37(2):183--233, 1999.

\bibitem[\protect\citeauthoryear{Karamanolakis \bgroup \em et al.\egroup
  }{2018}]{Karamanolakis2018}
Giannis Karamanolakis, Kevin~Raji Cherian, Ananth~Ravi Narayan, Jie Yuan,
  Da~Tang, and Tony Jebara.
\newblock Item recommendation with variational autoencoders and heterogeneous
  priors.
\newblock In {\em ACM DLRS}. Association for Computing Machinery, 2018.

\bibitem[\protect\citeauthoryear{Karatzoglou \bgroup \em et al.\egroup
  }{2010}]{Karatzoglou2010}
Alexandros Karatzoglou, Alexander Smola, and Markus Weimer.
\newblock Collaborative filtering on a budget.
\newblock {\em Journal of Machine Learning Research-Proceedings Track},
  9:389--396, 2010.

\bibitem[\protect\citeauthoryear{Kingma and Welling}{2014}]{kingma2014}
Diederik~P. Kingma and Max Welling.
\newblock Auto-encoding variational bayes.
\newblock {\em CoRR}, 2014.

\bibitem[\protect\citeauthoryear{Lee \bgroup \em et al.\egroup
  }{2017}]{Lee2017}
Wonsung Lee, Kyungwoo Song, and Il-Chul Moon.
\newblock Augmented variational autoencoders for collaborative filtering with
  auxiliary information.
\newblock In {\em ACM CIKM}, 2017.

\bibitem[\protect\citeauthoryear{Lian \bgroup \em et al.\egroup
  }{2017}]{Lian2017DiscreteCM}
Defu Lian, R.~Liu, Yong Ge, Kai Zheng, Xing Xie, and Longbing Cao.
\newblock Discrete content-aware matrix factorization.
\newblock In {\em ACM SIGKDD}, 2017.

\bibitem[\protect\citeauthoryear{Liang \bgroup \em et al.\egroup
  }{2018}]{Liang2018VAE}
Dawen Liang, Rahul~G. Krishnan, Matthew~D. Hoffman, and Tony Jebara.
\newblock Variational autoencoders for collaborative filtering.
\newblock In {\em WWW}, 2018.

\bibitem[\protect\citeauthoryear{Liu \bgroup \em et al.\egroup
  }{2019}]{Liu2019Compositional}
Chenghao Liu, Tao Lu, Xin Wang, Zhiyong Cheng, Jianling Sun, and Steven~CH Hoi.
\newblock Compositional coding for collaborative filtering.
\newblock In {\em ACM SIGIR}, 2019.

\bibitem[\protect\citeauthoryear{Liu \bgroup \em et al.\egroup
  }{2021}]{liu2021leveraging}
Weiming Liu, Jiajie Su, Chaochao Chen, and Xiaolin Zheng.
\newblock Leveraging distribution alignment via stein path for cross-domain
  cold-start recommendation.
\newblock In {\em NeurIPS}, volume~34, 2021.

\bibitem[\protect\citeauthoryear{Qi \bgroup \em et al.\egroup
  }{2021a}]{Qi2021Privacy}
Lianyong Qi, Chunhua Hu, Xuyun Zhang, Mohammad~R. Khosravi, Suraj Sharma,
  Shaoning Pang, and Tian Wang.
\newblock Privacy-aware data fusion and prediction with spatial-temporal
  context for smart city industrial environment.
\newblock {\em IEEE Transactions on Industrial Informatics}, 17(6):4159--4167,
  2021.

\bibitem[\protect\citeauthoryear{Qi \bgroup \em et al.\egroup }{2021b}]{Qi2021}
Lianyong Qi, Xiaokang Wang, Xiaolong Xu, Wanchun Dou, and Shancang Li.
\newblock Privacy-aware cross-platform service recommendation based on enhanced
  locality-sensitive hashing.
\newblock {\em IEEE Transactions on Network Science and Engineering},
  8(2):1145--1153, 2021.

\bibitem[\protect\citeauthoryear{Rendle \bgroup \em et al.\egroup
  }{2009}]{Rendle2009}
Steffen Rendle, Christoph Freudenthaler, Zeno Gantner, and Lars Schmidt-Thieme.
\newblock Bpr: Bayesian personalized ranking from implicit feedback.
\newblock In {\em ACM UAI}, 2009.

\bibitem[\protect\citeauthoryear{Rezende and Mohamed}{2015}]{Rezende2015}
Danilo~Jimenez Rezende and Shakir Mohamed.
\newblock Variational inference with normalizing flows.
\newblock In {\em ICML}, 2015.

\bibitem[\protect\citeauthoryear{Shan \bgroup \em et al.\egroup
  }{2018}]{Shan2018Recurrent}
Ying Shan, Jian jiao, Jie Zhu, and JC~Mao.
\newblock Recurrent binary embedding for gpu-enabled exhaustive retrieval from
  billion-scale semantic vectors.
\newblock In {\em ACM SIGKDD}, 2018.

\bibitem[\protect\citeauthoryear{Truong \bgroup \em et al.\egroup
  }{2021}]{Truong2021}
Quoc-Tuan Truong, Aghiles Salah, and Hady~W Lauw.
\newblock Bilateral variational autoencoder for collaborative filtering.
\newblock In {\em WSDM}, 2021.

\bibitem[\protect\citeauthoryear{Wang \bgroup \em et al.\egroup
  }{2018}]{Wang2018a}
Jingdong Wang, Ting Zhang, jingkuan song, Nicu Sebe, and Heng~Tao Shen.
\newblock A survey on learning to hash.
\newblock {\em IEEE Transactions on Pattern Analysis and Machine Intelligence},
  40(4):769--790, 2018.

\bibitem[\protect\citeauthoryear{Zhang \bgroup \em et al.\egroup
  }{2014}]{Zhang2014Preference}
Zhiwei Zhang, Qifan Wang, Lingyun Ruan, and Luo Si.
\newblock Preference preserving hashing for efficient recommendation.
\newblock In {\em ACM SIGIR}, 2014.

\bibitem[\protect\citeauthoryear{Zhang \bgroup \em et al.\egroup
  }{2016}]{Zhang2016DCF}
Hanwang Zhang, Fumin Shen, Wei Liu, Xiangnan He, Huanbo Luan, and Tat-Seng
  Chua.
\newblock Discrete collaborative filtering.
\newblock In {\em ACM SIGIR}, 2016.

\bibitem[\protect\citeauthoryear{Zhang \bgroup \em et al.\egroup
  }{2017a}]{Zhang2017DiscretePR}
Yan Zhang, Defu Lian, and Guowu Yang.
\newblock Discrete personalized ranking for fast collaborative filtering from
  implicit feedback.
\newblock In {\em IEEE AAAI}, 2017.

\bibitem[\protect\citeauthoryear{Zhang \bgroup \em et al.\egroup
  }{2017b}]{Zhang2017Dot}
Yan Zhang, Guowu Yang, Lin Hu, Hong Wen, and Jinsong Wu.
\newblock Dot-product based preference preserved hashing for fast collaborative
  filtering.
\newblock In {\em IEEE ICC}, 2017.

\bibitem[\protect\citeauthoryear{Zheng \bgroup \em et al.\egroup
  }{2020}]{Zheng2020EndtoEnd}
Minghang Zheng, Peng Gao, Xiaogang Wang, Hongsheng Li, and Hao Dong.
\newblock End-to-end object detection with adaptive clustering transformer.
\newblock {\em ArXiv}, abs/2011.09315, 2020.

\bibitem[\protect\citeauthoryear{Zhu \bgroup \em et al.\egroup
  }{2021}]{Zhu2021A}
Feng Zhu, Yan Wang, Jun Zhou, Chaochao Chen, Longfei Li, and Guanfeng Liu.
\newblock A unified framework for cross-domain and cross-system
  recommendations.
\newblock {\em CoRR}, 2021.

\end{thebibliography}

\end{document}